\documentclass[12pt]{amsart}
\usepackage[latin1]{inputenc}
\usepackage{amsmath}
\usepackage{amsthm}

\theoremstyle{plain}
 \newtheorem{thm}{Theorem}
 \newtheorem{lem}{Lemma} 
 \newtheorem{prop}{Proposition} 
 \newtheorem*{conj*}{Conjecture}
 \theoremstyle{definition}
 \newtheorem{defn}{Definition}
 \newtheorem{example}{Example}
 \newtheorem{remark}{Remark}

 \newtheorem{claim}{Claim}
 \newtheorem*{claim*}{Claim}

\newtheorem*{pironlemma}{Piron`s Lemma}

\newcommand{\calh}{\mathcal{H}} 

\newcommand{\resp}{{\it resp}}

\newcommand{\bfone}{\mathbf{1}}

\newcommand{\inner}[1]{\langle #1 \rangle}
 
\newcommand{\tr}{\text{tr}}

\newcommand{\EW}{\text{EW}}
\newcommand{\spann}[1]{\text{span}\{#1\}}

\begin{document}

\title{A Subjective Approach to Quantum Probability}
\author{Ehud Lehrer}\thanks{School of Mathematical Sciences, Tel aviv Universtity,
Tel Aviv 69978, Israel.\\ e-mails: \textsf{lehrer@post.tau.ac.il} ;
\textsf{gawain@post.tau.ac.il} }
\author{Eran Shmaya} \maketitle
\centerline{\today}
\begin{abstract}

A likelihood order is defined over linear subspaces of a finite
dimensional Hilbert space. It is shown that such an order that
satisfies some plausible axioms can be represented by a quantum
probability in two cases: pure state and uniform measure.

\end{abstract}

\section{Introduction}

According to the subjective approach probabilities are merely
degrees-of-belief of a rational agent. These degrees-of-belief
might be indicated by the agent's willingness to bet or take other
actions (see \cite{de finetti}). Savage (\cite{savage}) derives
both probabilities and utilities from rational preferences (i.e.,
that satisfy some putative properties) alone. Such preferences
induce, in particular, a preference order over events. That is, an
agent who holds rational preferences could indicate which of two
events is more likely, and moreover, this likelihood order is
transitive. Savage's first step is to derive a (finitely additive)
probability that represents the likelihood order.

In this paper we adopt a similar approach and apply it to the
quantum framework without going beyond probabilities. While
classical probability is defined over subsets (events) of a state
space, quantum probability is defined over subspaces of Hilbert
space. Furthermore, disjointness of the classical model is
replaced by orthogonality.

Formally, let $\calh$ be a separable Hilbert space. A quantum
probability measure $\mu$ over $\calh$ assigns a number between $0$
and $1$ to every closed subspace that satisfies $\mu(A\oplus
B)=\mu(A)+\mu(B)$ whenever $A \perp B$ and $\mu(\calh)=1$. Gleason's
Theorem (Gleason (1957)) states that, if $\dim(\calh)\geq 3$ every
quantum measure $\mu$ is induced by a self-adjoint nonnegative
operator $T$ with trace $1$ in the following way:
$\mu(A)=\text{tr}(\Pi_AT)$ for every subspace $A$, where $\Pi_A$ is
the orthogonal projection over $A$.

We assume the existence of a likelihood order $\preceq$ over
subspaces of a given finite-dimensional Hilbert space. The statement
`$A$ is less likely than  $B$ in one's eyes' could be understood
operationally: one would prefer betting that $B$ occurs  than that
$A$ occurs (in the corresponding physical measurements).

We say that the likelihood order $\preceq$ can be \emph{represented
by a quantum probability} $\mu$ if $A \preceq B$ if and only if
$\mu(A) \le \mu(B).$ The goal of the line of research presented here
is to find  plausible properties (axioms, in the jargon of decision
theory), preferably  rationality-motivated, that ensure that
$\preceq$ is  representable by a quantum probability. Such a
representation would mean that the agent acts as if he has
quantitative degrees-of-belief that obey the rules of quantum
probability.

Throughout, it is assumed that $\preceq$  possesses three
properties. The first is that every subspace is more likely than
the zero-dimensional one. The second is that  a subspace $B$ is
more likely than $A$ if and only if  $B+C$ is more likely than
$A+C$, whenever $C\perp A$ and $C\perp B$. That is, adding or
deleting a subspace which is orthogonal to both $A$ and $B$ would
preserve the likelihood ratio.

The classical counterpart of the third property is a consequence of
the second. However, in the quantum model it has to be explicitly
assumed. It states that if $B$ is more likely than $A$, then the
orthogonal complement of $B$ is less likely than that of $A$.

Savage (\cite{savage}) also assumes these three axioms but, in order
to obtain a representation by a measure, he needs an additional,
less motivated, property that concerns with the richness of the
state space. This one dictates that the state space could be split
into mutually disjoint arbitrarily small (with respect to the
likelihood order) subsets. The lack of a quantum counterpart (in the
case of a finite-dimensional Hilbert space) of such an Archimedean
property makes our study completely different from that of Savage.

Our main results refer to likelihood orders that can be represented
by two types of quantum measures. The first is the most important
from a physical point of view. The probabilities of this type are
called \emph{pure states} and are of the form
$\mu(A)=\|\Pi_A(p)\|^2$ for some unit vector $p\in \calh$. That is,
the probability of a subspace $A$ is the length squared of the
projection of the vector $p$. By Gleason's Theorem these measures
are the extreme points of the convex set of all quantum
probabilities. We characterize the likelihood orders that can be
represented by a quantum measure.

The second main result characterizes the likelihood orders that can
be represented by the uniform distribution, defined by
$\mu(A)=\frac{\dim(A)}{\dim(\calh)}$. This is the only quantum
measure that obtains a discrete set of values.

\bigskip

Subjective analysis of quantum probability has been treated in the
literature by several authors. Deutsch (\cite{deutsch},
\cite{prodeutsch}) assumes that an agent assigns a value to any
possible outcome of any possible measurement.
Deutsch's analysis hinges heavily on what he calls the `principle of
substitutibility', which is similar to the Independence Axiom of von
Neumann and Morgenstern (\cite{vN-M}). Barnum et al.
(\cite{antideutsch}) criticized Deutsch's argument and showed that
his proof relies on a tacit symmetry assumption.
Wallace (\cite{prodeutsch}) followed the line of Deutsch
(\cite{deutsch}) and tried to make his assumptions more plausible.
Gyntelberg and Hansen (\cite{gyntelberg}) applied a general
event-lattice theory (with axioms that resemble those of von Neumann
and Morgenstern) to a similar framework.

Pitowsky (\cite{pitowsky}) assumed that for every possible
measurement the agent has a certain probability over the
corresponding outcomes. From some natural axioms he derives the
probabilistic structure over quantum mechanics. Caves et al.
(\cite{caves}) assume that the agent has degrees-of-belief that
determine the odds under which he is willing to take a bet. Under
the assumption that the agent cannot be attacked by a Dutch book,
and an assumption about `maximal information', they showed that
these degrees-of-belief must be given by a pure state.

The main difference between the aforementioned approaches and ours
is that we do not assume that the agent has quantitative
assignments: neither probabilities (i.e., numerical
degrees-of-belief) to subspaces nor values to games or lotteries.
Rather, the primitive of our model is a qualitative belief given by
the likelihood order.

\bigskip

The paper is structured as follows. The next section characterizes
the likelihood orders that admit a quantum probability
representation in terms of continuity and a duality-like condition,
called the cancelation condition. Section \ref{sec-axioms}
introduces the main axioms.  Sections \ref{sec-pure} and
\ref{sec-uniform} are devoted to the main results: representation by
a pure state and by a uniform distribution. Section
\ref{sec-counter} provides an example of a likelihood order that
satisfies the main axioms except for continuity, and cannot be
represented by a quantum measure. The paper is concluded  with
Section \ref{sec-final} that provides some final comments and open
problems.

\section{The Cancelation Condition and Continuity}
Let  $\calh$ be  a finite dimensional Hilbert space and let
$\preceq$ be a weak order over linear subspaces of  $\calh$, that is
$\preceq$ is reflexive (i.e., for every $A$, $A\preceq A$),
transitive (i.e., for every $A,B,C$, if $A\preceq B$ and $B\preceq
C$ then $A\preceq C$) and complete (i.e., for every $A,B$, $A\preceq
B$ or $B\preceq A$ or both). We call $\preceq$ the \emph{likelihood}
order, and when $A \preceq B$ we say that $B$ is \emph{more likely
than} $A$. Denote by $\sim$ the equivalence relation induced by
$\preceq$ (i.e., $A\sim B$ if $A\preceq B$ and $B\preceq A$) and by
$\prec$ the corresponding strict order (i.e., $A\prec B$ if
$A\preceq B$ and $B\not\preceq A$.)

\subsection{The cancelation condition}
Cancelation condition (see, for example~\cite{fishburn}) is a
well-known property of a weak order in the classical framework:
\begin{prop}
\label{canclassical} Let $\preceq$ be a weak order over subsets of a
finite set $\Omega$. For $A\subseteq\Omega$, denote by $\bfone_A$
the indicator function of $A$. Then there exists an additive
probability measure $\mu$ over $\Omega$ such that $A\preceq
B\leftrightarrow \mu(A)\leq \mu(B)$ for every $A,B\subseteq \Omega$
if and only if the following conditions hold:
\begin{enumerate}\item For every $A\subseteq\Omega$, $\Phi\preceq
A$.\item $\Phi\prec \Omega$.\item For every $n$, if
$A_1,\dots,A_n,B_1,\dots,B_n$ are subsets of $\Omega$ such that
$\sum_{i=1}^n\bfone_{A_i}=\sum_{i=1}^n\bfone_{B_i}$ and $A_i\preceq
B_i$ for every $i$, then $A_i\sim B_i$ for every $i$.\end{enumerate}
\end{prop}

\begin{defn}
Let $\mu$ be a quantum probability measure over $\calh$. We say that
$\mu$ \emph{represents} $\preceq$ if $A\preceq B\leftrightarrow
\mu(A)\leq \mu(B)$ for every two subspaces $A,B$ of $\calh$.
\end{defn}

In the quantum framework, orthogonal projection will replace the
indicator functions that appear in Lemma ~\ref{canclassical}.




\begin{defn}
The likelihood order $\preceq$ satisfies the \emph{cancelation
condition} if, for every $2n$ subspaces of $\calh$,
$A_1,\dots,A_n,B_1,\dots,B_n$, and $n$ positive numbers
$\alpha_i$,
$i=1,...,n$,
if 
 $\sum_{i=1}^n\alpha_i\Pi_{A_i}= \sum_{i=1}^n\alpha_i\Pi_{B_i}$ and
$A_i\preceq B_i$ for every $i=1,...,n$, then  $A_i\sim B_i$ for
every $i=1,...,n$.
\end{defn}

\subsection{Continuity}
The cancelation condition by itself is not sufficient to ensure the
existence of a representative measure (see Example ~\ref{lex}
below.) Similar examples appear in the classical framework, when one
tries to extend Proposition~\ref{canclassical} to an infinite
$\Omega$. In the current framework, in order to obtain a
characterization of the likelihood orders that can be represented by
a quantum measure, we need the additional assumption that $\preceq$
is continuous w.r.t. the natural topology over subspaces.

Let $A$  be subspace and  $r$ be a positive number. Denote by $U$
the unit ball, $U=\{x;~ \|x\| \le 1\}.$ By $N_r(A)$ we denote the
$r$-neighborhood of $A$ restricted to the unit ball, $\cup\{ V(x,
r); ~x \in A ~{\rm and}~ x\in U\}, $ where $V(x, r)$ is the ball of
radius $r$ around $x$. For two subspaces $A$ and $B$ we denote
$\delta_*(A,B) = \inf\{r>0 :~ A \cap U\subset N_r(B)\}$ and $
\delta^*(A,B)=\inf\{r>0 :~ B\cap U\subset N_r(A)\}$. The Hausdorff
metric is defined as

 $$\delta(A,B) = \max\{\delta^*(A,B),\delta_*(A,B)\}.$$

\begin{defn}
The likelihood order $\preceq$ is \emph{lower semi-continuous} if,
for every subspace $B$, the set of the subspaces $A$ such that
$A\prec B$ is open with respect to $\delta$.
\end{defn}


\begin{thm}
\label{canquantum} Let $\preceq$ be a likelihood order. There
exists a quantum probability measure that represents $\preceq$ if
and only if the following conditions are satisfied:
\begin{enumerate}\item $\{0\}\preceq A$ For every subspace $A$ of $\calh$;\item $\{0\}\prec
\calh$;
\item $\preceq$ is lower semi-continuous;\item $\preceq$ satisfies the cancelation
condition.\end{enumerate}
\end{thm}

\begin{proof}
Assume first that $\preceq$ is represented by a quantum probability
$\mu$. Then by Gleason's Theorem there exists a nonnegative operator
$T$ with trace $1$ such that $\mu(A)=\tr(\Pi_A T)$ for every
subspace $A$ of $V$. In particular, the function $A\mapsto \mu(A)$
is continuous and therefore the order that it represents is lower
semi-continuous. As for the cancelation condition, let
$A_1,\dots,A_n,B_1,\dots,B_n$ be subspaces such that
$\sum_{i=1}^n\alpha_i\Pi_{A_i}= \sum_{i=1}^n\alpha_i \Pi_{B_i}$ and
$A_i\preceq B_i$, where $\alpha_i$, $i=1,...,n$, are positive
numbers 
It follows that
\begin{multline*} \sum_i \alpha_i\mu(A_i)=\sum_i\alpha_i
\tr(\Pi_{A_i}T)=\tr\bigl((\sum_i\alpha_i \Pi_{A_i})T\bigr)=\\
 \tr\bigl((\sum_i \alpha_i \Pi_{B_i})T\bigr)=
 \sum_i\alpha_i \tr(\Pi_{B_i}T)=\sum_i \alpha_i\mu(B_i).\end{multline*}
Since $\mu(A_i)\leq \mu(B_i)$ for every $i$, it follows that
$\mu(A_i)=\mu(B_i)$, which implies $A_i\sim B_i$.

Consider the finite dimensional Hilbert space of Hermitian
operators over $\calh$ with the inner product of two Hermitian
operators $S$ and $T$ being $\tr (ST)$. Denote
$\mathcal{C}=\text{Conv}\{\Pi_B-\Pi_A;~A\prec B\}$ and
$\mathcal{D}=\spann{\Pi_B-\Pi_A;A\sim B}$. $\mathcal{C}$ is a
convex set and $\mathcal{D}$ is a linear subspace. From the
cancelation condition it follows that $\mathcal{C}$ and
$\mathcal{D}$ are disjoint.

The separation theorem (see, \cite{Rockefeller}) ensures that there
is a non-zero linear function,  represented in this case by the
Hermitian operator $T$, such that $\tr(DT)=0$ for every $D \in
\mathcal{D}$ (since $\mathcal{D}$ is a subspace) and $\tr(CT)\geq 0$
for every $C\in\mathcal{C}$.

Since $\tr(DT)=0$ for $D\in\mathcal{D}$, it follows that
$\tr(AT)=\tr(BT)$ if $A\sim B$. Let $A\prec B$. By the definition of
$\mathcal{C}$, $\Pi_{A}- \Pi_{B}\in \mathcal{C}$. Due to lower
semi-continuity, for $B'$ sufficiently close to $B$, $A\prec B'$ and
therefore $\Pi_{A}- \Pi_{B'} \in \mathcal{C}$. However, the set
$\text{Conv}\{\Pi_{A}- \Pi_{B'}\}$, where $B'$ is sufficiently close
to $B$, contains $\Pi_{A}- \Pi_{B}$ as an interior point. Therefore,
$\tr((\Pi_{A}- \Pi_{B})T)$ is strictly positive. We conclude that
$T$ represents $\preceq$.

Finally, for every $A$, $\tr(\Pi_A T)\geq 0$ since $\{0\}\preceq A$.
Therefore, $T$ is positive semidefinite. Moreover, since $\{0\}\prec
\calh$, it follows that $\tr(T)>0$. Define $T'=\frac{T}{\tr(T)}$. We
obtain that $\preceq$ is represented by $\mu(A)=\tr(T' A)$, $T'$ is
positive semidefinite and $\tr(T')=1$, as desired. \end{proof}

\bigskip
The cancelation condition (even in the classical framework) is
difficult to justify.  It is desirable to derive a probability
representation of a likelihood order over linear subspaces from more
plausible assumptions.

\section{de-Finetti's and Other Axioms}\label{sec-axioms}

The most natural condition is de-Finetti's. When applied to
classical probability it states that $C$ is disjoint of $A \cup B$,
then $B$ is preferred to $A$ iff $B\cup C$ is preferred to $A\cup
C$. In the quantum framework it takes following form:

\bigskip

  \textbf{de-Finetti's Axiom}: For every linear
subspaces $A,B,C$ of $\calh$, if $A\perp C$ and $B\perp C$, then
$A\preceq B$ iff $A+C\preceq B+C$.

\bigskip

In the classical framework it easily follows from
\textbf{de-Finetti's axiom} that if $A\preceq B$ then $B^c\preceq
A^c$. In the quantum framework, we need to require it explicitly.

\bigskip

\textbf{Negation}: For every two linear subspaces $A,B$ of $\calh$,
if $A\preceq B$ then $B^\perp \preceq A^\perp$.

\bigskip

As illustrated by the following example there might be weak orders
that satisfy \textbf{de-Finetti's axiom} and not \textbf{Negation}.

\begin{example} Let  $\calh$ be $\mathbb{R}^2$. Any monotonic (w.r.t set inclusion) weak
order on  $\calh$ satisfies \textbf{de-Finetti's axiom} but it might
not satisfy \textbf{Negation}. Let $\succeq'$  be such a weak order.
As for a higher dimensional Hilbert space, let $\calh$ be
$\mathbb{R}^3$. Let $p$ be the northern pole of the unit ball, $E$
be the equator, and define $\mu(u)=|\inner{p,u}|^2$ for any unit
vector $u$.

Define $\preceq$ as follows: If $A$ and $B$ are two subspaces of
different dimensions, then $A\succ B$ if the dimension of $A$ is
greater than that of $B$. If $A=\spann{u}$ and $B=\spann{v}$, where
$u$ and $v$ are unit vectors, then $A\succeq B$ either when $u\not
\in E$ and $\mu(u) \ge \mu(v)$ or when $u,v \in E$ and $u \succeq'
v$. Finally, if $A$ and $B$ are two-dimensional subspaces, then
$A\succeq B$ either when $p \not \in B$ and $\|\Pi_A(p)\|^2 >
\|\Pi_B(p)\| ^2$ or when $p   \in A \cap B$ and $A\cap E \succeq'
B\cap E$. The weak order $\preceq$ preserves \textbf{de-Finetti's
axiom} but since $\succeq'$ does not preserve \textbf{Negation} on
$E$, so $\succeq$ does not on $\calh$.
\end{example}

\smallskip

We will also need the obvious assumption that any subspace is as
preferred as the zero-dimensional one. Formally,

\bigskip

\textbf{Monotonicity}: For every subspace $A$ of $\calh$,
$\{0\}\preceq A$.

\bigskip

Note that\textbf{ Monotonicity} and \textbf{de-Finetti's axiom}
together imply that if $A\subseteq B$ then $A\preceq B$ for every
pair $A,B$ of subspaces. Thus, $\preceq $ is monotonic with respect
to set inclusion.
\bigskip

In the sequel, we will say that a weak order $\preceq$ satisfies the
\emph{standard assumptions} if it satisfies \textbf{de-Finetti's
axiom}, \textbf{Negation} and \textbf{Monotonicity}.

Do the standard assumptions guarantee that $\preceq$ can be
represented by a measure? The following is a counterexample.

\begin{example} \label{lex} Let $\mu_1,\mu_2$  be two
different quantum probability measures over $\calh$ and define the
\emph{lexicographic order} induced by  $\mu_1$ and $\mu_2$ as
follows. $A\preceq B$ if either $\mu_1(A)<\mu_1(B)$ or
$\mu_1(A)=\mu_1(B)$ and $\mu_2(A)<\mu_2(B)$. Then, $\preceq$
satisfies the standard assumptions. Furthermore, it satisfies the
cancelation condition.\end{example}

The lexicographic order cannot be represented by a measure since it
lacks the following property:
\bigskip

\textbf{Separability}: There is a countable set of subspaces,
$\mathcal{A}$,  such that for any two subspaces $B$ and $C$ such
that $B \prec C,$ there is $A \in \mathcal{A}$ that satisfies $B
\preceq A \preceq  C.$

\bigskip
As indicated by Debreu (\cite{debreu}), \textbf{Separability} is
necessary for  $\preceq$ in order to be represented by a real
function (not necessarily a measure).

\section{Pure States in $\mathbb{R}^3$}\label{sec-pure}
\label{purestate} The most important probabilities from the physical
point of view are those of the form $\mu(A)=\|\Pi_A(p)\|^2$ for some
unit vector $p\in A$. These distributions are sometimes called
\emph{pure states}. It follows from Gleason's Theorem that pure
states are the extreme points of the convex set of all quantum
probabilities.

It is clear that if $\mu$ is a pure state and $\preceq$ is the
induced likelihood order, then the one-dimensional subspace spanned
by $p$ is equivalent (under $\sim$) to $\calh$. In this section we
prove the inverse statement. We say that $\preceq$ is
\emph{non-trivial} if there exists a subspace $A$ such that $\{0\}
\prec A.  $

\begin{thm}\label{purethm}
Let $\preceq$ be a  weak order over subspaces of a finite
dimensional real-Hilbert space that satisfies the standard
assumptions and \textbf{Separability}.  Assume that there exists a
one-dimensional subspace $P$ such that $P\sim\calh$. Let $p$ be a
unit vector in $P$. If $\preceq$ is non-trivial, then $\preceq$ is
represented by the pure state $\mu(A)=\|\Pi_A(p)\|^2$.
\end{thm}

\begin{proof}
Let $E$ be the orthogonal complement of $P=\spann{p}$. By
\textbf{Negation}, since $P\sim\calh$ it follows that $E\sim\{0\}$.
Let $A$ be a subspace of $\calh$, $A_0$ be $A_0=A\cap E$ and $A_1$
be the one-dimensional subspace of $\calh$ that is spanned by
$\Pi_A(p)$ (the orthogonal projection of $p$ over $A$.) Then, since
$A_0\subset E$, it follows from \textbf{Monotonicity} that $A_0\sim
0$. Since $A_1\perp A_0$ and $A=A_1+A_0$, it follows from
\textbf{de-Finetti's axiom} that $A\sim A_1$. Thus, the likelihood
order  $\preceq$ is determined by its restriction to one-dimensional
subspaces. Moreover, since $\mu(A)=\mu(A_1)$, it is sufficient to
prove that the likelihood order over one-dimensional subspaces is
represented by $\mu$. Slightly abusing notation, we will identify a
unit vector $u$ in $\calh$ with the one-dimensional subspace spanned
by $u$. With this convention, for every unit vector $u$,
$\mu(u)=|\inner{p,u}|^2$.

Let $u,v$ be two unit vectors. We need to show that $u\preceq v$ iff
$|\inner{p,u}|^2\leq |\inner{p,v}|^2$. By looking at the
three-dimensional space $\calh_{uv}$ spanned by $p,u,v$, with its
two-dimensional subspace $\calh_{uv}\cap E$ we can assume w.l.o.g.
that $\dim{\calh}=3$. In this case the theorem will follow directly
from the following proposition. Let $S^2$ be the unit sphere in
$\mathbb{R}^3$.  We say that $ \preceq$ is \emph{uniform} if all the
one-dimensional subspaces are equivalent.
\begin{prop}
\label{pure_r3} Let $\preceq$ be a weak order over $\mathbb{R}^3$
that satisfies the standard assumptions, and such that the
restriction of $\preceq$ to one-dimensional subspaces is separable.
Assume that $\preceq$ is not uniform and it attains its minimum over
$S^2$ at $m$. Furthermore, assume that there exists a
two-dimensional subspace $E$ such that $m\sim u$ for every $u\in E$.
Let $p\in S^2$ such that $p\perp E$. Then, for every pair $u,v\in
S^2$, $u\preceq v$ iff $|\inner{p,u}|^2\leq |\inner{p,v}|^2$.
\end{prop}
\emph{Proof of Proposition~\ref{pure_r3}.} The proof of the
proposition is broken into five claims.  As usual, we identify
elements of $S^2$ with their corresponding one-dimensional
subspaces.
\begin{claim}\label{claim_0}
Let $q,r\in S^2$ be such that $q\preceq r$. If $q'$ and $r'$ are,
respectively,  the orthogonal complements of $q$ and $r$ in the
plane $\spann{q,r}$, then $r'\leq q'$.
\end{claim}
\begin{proof} Let $n\in S^2$ such that $n\perp\spann{q,r}$. Then
$q^\perp = \spann{n,q'}$ and $r^\perp=\spann{n,r'}$. By
\textbf{Negation}, $r^\perp\preceq q^\perp$. By \textbf{de-Finetti's
axiom}, $r'\preceq q'$.
\end{proof}
\begin{claim}
\label{claim_01} Let $u_1,u_2\in S^2$ be orthogonal vectors such
that $u_1\sim u_2\sim m$. If $u\in\spann{u_1,u_2}$, then $u\sim m$.
\end{claim}
\begin{proof}
Note that $u_2$ is the orthogonal complement of $u_1$ in
$\spann{u_1,u_2}$. Let $u'$ be the orthogonal complement of $u$ in
$\spann{u_1,u_2}$. Then $u_1\sim m\preceq u'$. By
Claim~\ref{claim_0}, $u\preceq u_2\sim m$. Since $m\preceq u$, it
follows that $u\sim m$.
\end{proof}
\begin{claim}\label{claim_02}
Assume that there exists an orthogonal triple $u_1,u_2,u_3$ such
that $m\sim u_1\sim u_2\sim u_3$. Then, $ \preceq$ is uniform.
\end{claim}
\begin{proof}Let $v\in S^2$. Then there exists $u\in\spann{u_1,u_2}$
such that $v\in\spann{u,u_3}$. By Claim~\ref{claim_01}, $u\sim m$.
But $u\perp u_3$ and therefore again by Claim~\ref{claim_01}, $v\sim
m$.
\end{proof}
For $q\in S^2$ we denote by $\theta(p,q)$ the angle between $p$ and
$q$. Thus $0\leq\theta(p,q)\leq \pi$ and
$\cos\theta(p,q)=\inner{p,q}$. Let $N_p = \{q \in S^2 |~ 0 <
\theta(p,q) < \pi/2\}$ be the \emph{northern hemisphere} relative to
$p$, and  $E_p = \{q \in S^2 | \theta(p,q)=\pi/2\}=E\cap S^2$ be the
\emph{equator} relative to $p$. Let $q\in N_p$. Among the great
circles which pass through $q$ there is a unique one that intersects
$E_p$ in vector $x$ orthogonal to $q$. We follow Gleason (1957) and
denote this circle by $\EW(q)$. Note that $q$ is the northern most
point in $\EW(q)$ and that $\EW(q)$ is tangent to the latitude
circle of $q$. We will need the following lemma, that appears in
(\cite{piron}) (see also \cite{cooke}).
\begin{pironlemma} Let $q,r \in N_p$ such that $\theta(p,q) < \theta(p,r)$; then there exists a
finite sequence $q=q_0,q_1,\dots,q_n=r$ of points in $N_p$ such that $q_{i+1}\in\EW (q_i)$.
\end{pironlemma}
\begin{claim}
\label{claim_1} Under the assumption of Theorem~\ref{purethm}, if
$q,r \in N_p$ and $\mu(r) < \mu(q)$, then $r \prec q$.
\end{claim}
\begin{proof}
Note that $\mu(q)>\mu(r)$ iff $\theta(p,q)<\theta(p,r)$.

Let $q \in N_p$ and $q_1 \in \EW(q)$. Let $q_1'\in EW(q)$ be the
orthogonal complement of $q_1$ in the plane of $\EW(q)$, and $q'\in
E_p$ be the orthogonal complement of $q$ in $\EW(q)$. Since $q'\sim
m\preceq q_1'$, it follows from Claim~\ref{claim_0} that $q_1\preceq
q$. Moreover, $q_1 \sim q$ only if $q_1' \sim m$. By induction it
follows from Piron's Lemma that $r \preceq q$. Furthermore, $r \sim
q$ only if there exists  $z \in N_p$ such that $z\sim m$. We prove
that in this case $\preceq$ is uniform, which is excluded by
assumption. This will complete the proof.

Note that for every $y$ such that $\theta(p,z) < \theta(p,y) < \pi -
\theta(p,z)$, $m \preceq y \preceq z \sim m$. Thus, all the vectors
in the band below $z$ are equivalent to $m$. We now show that one
can find another point $p'$, such that $x'\sim m$ for every $x'\in
E_{p'}$ and $\theta(p',z) =  \frac{1}{2} \theta(p,z)$,  and thus
obtaining a wider band. By iterating this argument one can get wider
and wider bands until one obtains a band that is wide enough to
contain three orthogonal vectors. By Claim~\ref{claim_02}, it would
imply that $\preceq$ is uniform.

Let $p'$ be point in $\spann{p,z}$ for which
$\theta(p,p')=\theta(p',z)= \frac{1}{2} \theta(p,z)$. It follows
that, for every $x' \in E_{p'}$, $\frac{3}{2}\theta(p,z) <
\theta(p,x') < \pi - \frac{3}{2}\theta(p,z)$. Thus, $E_{p'}$ is
entirely contained in the band defined by $p$ and $z$ and therefore
$E_{p'}\sim m$.
\end{proof}
\begin{claim}
\label{claim_2} If $q,r \in N_p$ and $\mu(q) = \mu(r)$, then $q\sim
r$.
\end{claim}
\begin{proof}
We know from Claim~\ref{claim_1} that for $q,r \in N_p$ such that
$\mu(q) < \mu(r)$, $r \prec q$. Now suppose that there exist, for
some $\alpha$, $0 < \alpha < \frac{\pi}{2}$, vectors $q_0,r_0\in
N_p$ such that $q_0\prec r_0$ and $\mu(q_0)=\mu(r_0)=\alpha$. Let
$Q=\{q\in S^2; ~\mu(q)=\alpha,q\prec r_0\}$ and $R=\{r\in S^2;
~\mu(r)=\alpha,q_0\prec r\}$. Then at least one of the sets $Q,R$
must be uncountable. Assume w.l.o.g. that $Q$ is uncountable. For
every $q\in Q$, let $q',r'$ be the orthogonal complements of $q,r_0$
\resp. in $\spann{q,r_0}$. It follows from Claim~\ref{claim_0} that
$r'\prec q'$. Notice moreover, that
$\mu(q')=\mu(r')=1-\alpha-\mu(n(q))$ where $n(q)\in S^2$ is
orthogonal to $\spann{q,r_0}$. Since
$\theta(p,n(q))=\cos\sqrt{\mu(n(q))}$ increases as $q$ approaches
$r_0$ along the latitude circle of $r_0$, we get uncountable set of
pairs $(r',q')$ such that $\mu(r')=\mu(q')$, but $r'\prec q'$ with
different values of $\mu$ for different pairs. This, together with
Claim~\ref{claim_1} contradicts separability.
\end{proof}

>From Claims \ref{claim_1} and \ref{claim_2} it follows that
$\mu(q)\leq\mu(r)$ iff $q\preceq r$, and therefore the proof of
Proposition \ref{pure_r3} is complete.

\medskip

Back to the proof of Theorem \ref{purethm}. By assumption, $\preceq$
is non-trivial. Therefore, there exists a subspace $A$ which is
strictly more likely than $\{0\}$. Suppose that $A$ is spanned by
the orthogonal vectors $u_1,...,u_k$.

\begin{claim}
\label{claim_non-trivial} At least one $u_i$ is strictly more likely
than  $\{0\}$.
\end{claim}
\begin{proof} Otherwise, $u_i \sim \{0\}$ for every $i=1,...k.$
By  \textbf{de-Finetti's axiom} $\spann{u_1,u_2}\sim \spann{u_2}
\sim \{0\}$. By successively adding the $u_i$'s and by using
\textbf{de-Finetti's axiom} one obtains that $A\sim \{0\}$, in
contradiction with the assumption.\end{proof}

By Claim \ref{claim_non-trivial} we can assume that there is a
vector $x \in S^2$ such that $x \succ \{0\}$. Let $y\in \spann{x,p}
\cap E$. Since $y \in E$, $y \sim \{0\}$. Furthermore, $y\perp p$.
Let $x'$ be the orthogonal complement of $x$ in $\spann{x,p} $.
Since $x' \succ y$, by  Claim~\ref{claim_0}  $x \preceq  p$. As $
\preceq  $ is an order,  $\{0\} \prec x \preceq  p$, and thus,
$\{0\} \prec  p$. This implies that
 $\preceq$, when restricted to $\calh_{uv}$, is
not uniform, as assumed by Proposition  \ref{pure_r3}. This enables
us to use this proposition in order to complete the proof of Theorem
\ref{purethm}.
\end{proof}

\begin{remark}
No sort of continuity is assumed in  Theorem \ref{purethm}.
Nevertheless, $\preceq$ is represented by a measure and is therefore
continuous.
\end{remark}

\section{The uniform measure}\label{sec-uniform} The only quantum probability measures over
a finite dimensional Hilbert space $\calh$ which receives discrete
values is given by the \emph{uniform measure},
$\mu(A)=\frac{\dim(A)}{\dim(\calh)}$. It turns out that this is the
case characterized by the property that all one-dimensional
subspaces are equally likely. Formally,
\begin{prop}
\label{trivial-uniform} Let $\preceq$ be a weak order over subspaces
of a finite dimensional Hilbert space that satisfies
\textbf{de-Finetti's axiom}. If all one-dimensional subspaces are
equivalent, then either $\preceq$ is trivial (i.e. $\{0\}\sim A$ for
every subspace $A$ of $\calh$) or $\preceq$ is represented by the
uniform measure.
\end{prop}
\begin{proof}
Assume that every one-dimensional subspace is equivalent to some
one-dimensional subspace, say, $m$. If $A_1,A_2$ are two-dimensional
such that $A_1\cap A_2$ is one-dimensional, we can assume that
$A_1=\spann{a_0,a_1}$ and $A_2=\spann{a_0,a_2}$ where $a_0\perp a_1$
and $a_0\perp a_2$. Since $a_1\sim a_2\sim m$ we get, by
\textbf{de-Finetti's axiom}, that $A_1\sim A_2$. If $A_1\cap
A_2=\{0\}$, we can find a two-dimensional subspace $A'$ such that
$A_1\cap A'$ and $A_2\cap A'$ are one-dimensional. Therefore
$A_1\sim A'\sim A_2$. Thus every two-dimensional subspaces are
equivalent. By a similar argument, two subspaces of the same
dimension are equivalent.

Finally, if $\{0\}\sim m$, it follows by \textbf{de-Finetti's axiom}
that $\{0\}\sim \calh$. If $0\prec m$, then again by
\textbf{de-Finetti's axiom}, if $A\perp m$ and $A'=A+m$ then $A\prec
A'$. Using the equivalence of two subspaces with the same dimension,
it follows that if $\dim(A')=\dim(A)+1$, then $A\prec A'$ and
therefore $\preceq$ is represented by
$\mu(A)=\frac{\dim(A)}{\dim(\calh)}$.
\end{proof}

The following nontrivial fact about quantum probabilities follows
from Gleason's Theorem:
\begin{prop}
Let $\calh$ be a finite-dimensional Hilbert space and $\mu$ be a
quantum probability over $\calh$. Assume that there exist
one-dimensional subspaces (not necessarily orthogonal)
$u_1,\dots,u_n$ of $\calh$ such that $\calh=u_1+\dots+u_n$ and
$\mu(u_1)=\dots=\mu(u_n)\leq\mu(x)$ for every one-dimensional
subspace $x$ of $\calh$. Then, $\mu$ is the uniform measure.
\end{prop}
We show that this proposition is a consequence of the standard
assumptions, with the additional assumption that $\preceq$ is
continuous over one-dimensional subspaces.

\begin{defn}
The likelihood order $\preceq$ is \emph{continuous  over
one-dimensional subspaces} if for every unit vector $v$ the sets
$\{u;~ u \prec v \}$ and $\{u;~v\prec u\}$ are  open.
\end{defn}
We note that if $\preceq$ is continuous over one-dimensional
subspaces then its restriction to one-dimensional subspaces is also
separable. Indeed, let $D\subseteq S^2$ be a countable dense set
w.r.t. the Euclidean topology of $S^2$. For every $u,v\in S^2$ such
that $u\prec v$, let $U=\{u'\in S^2|u\prec u'\}$ and $V=\{v'\in
S^2|v'\prec v\}$.  Since $S^2=U\cup V$ and $S^2$ is connected,
$U\cap V\neq\phi$. As $D$ is dense, there exists $d\in D$ such that
$d\in U\cap V$. Thus, $u\prec d\prec v$.

\smallskip

We state the result in $\mathbb{R}^3$. It can easily be extended to
every finite-dimensional Hilbert space.

\begin{thm}
\label{triple} Let $\preceq$ be a weak order over $\mathbb{R}^3$
that satisfies the standard assumptions. Assume that $\preceq$ is
continuous over one-dimensional subspaces. If $u_1, u_2 , u_3$ is a
basis (not necessarily orthogonal) that satisfies $u_1\sim u_2 \sim
u_3 \sim m$, where $m$ is a minimum of $\preceq$, then $x \sim m$
for every $x \in S^2$.
\end{thm}

The theorem is proved in a few steps. Denote by $M$ a maximum of
$\preceq$.

\begin{claim}\label{mMtags}
Let $u,v\in S^2$ such that $u\sim m$ and $v\sim M$. Let $u',v'$ be
the orthogonal complements of $u,v$ in $\spann{u,v}$, respectively.
Then, $u'\sim M$ and $v'\sim m$.
\end{claim}
\begin{proof}
Since $m\sim u\preceq v'$ it follows from Claim~\ref{claim_0} that
$v\preceq u'$. But $v\sim M$ and $M$ is a maximum. Therefore $u'\sim
M$. By a similar argument $v'\sim m$. \end{proof}
\begin{claim}
\label{precispure} If $u_1\perp u_2 \in S^2$ and $m \sim u_1 \sim
u_2$, then either all one-dimensional subspaces are equivalent or
$\preceq$ is represented by a pure state.
\end{claim}

\begin{proof}
Let $E=\spann{u_1,u_2}$. By Claim~\ref{claim_01}, $u\sim m$ for
every $u\in E$. By Proposition~\ref{pure_r3}, either $\preceq$ is
trivial or $\preceq$ is represented by a pure state.
\end{proof}
\begin{claim}
\label{claim mmp} If $u_1 \neq \pm u_2 \in S^2$, $M\sim u_1 \sim
u_2$ and $p \perp u_1,u_2$, then $p \sim m$.
\end{claim}
\begin{proof}
Let $x \in N_p$ be such that $x \sim m$ and $\theta(p,x)$ is
minimal. If $x \neq p$ then $x$ cannot be orthogonal to both $u_1$
and $u_2$. Assume therefore w.l.o.g. that $\langle x,u_1\rangle\neq
0$. Let $u_1'$ be the orthogonal complement of $u_1$ in the plane
$\spann{x,u_1}$. By Claim~\ref{mMtags}, $u_1'\sim m$. Moreover
$\theta(p,u_1')<\theta(p,x)$ since $x \in EW(u_1')$. This
contradicts  the choice of $x$. It therefore follows that $x = p$,
meaning that  $p \sim m$.
\end{proof}
\begin{claim}
If $m$ and $M$ are any minimal and maximal elements in $S^2$ and $m
\prec M$ (i.e., $\preceq$ is not trivial), then $m \perp M$.
\end{claim}
\begin{proof}
Assume the contrary. Let $a$ be the orthogonal complement of $m$ in
$\text{span}(m,M)$. By Claim~\ref{mMtags}, $a\sim M$. Let $p$
satisfy $p\perp M$ and $p \perp a$. By Claim~\ref{claim mmp}, $p
\sim m$. However, since $m \in \text{span}(M,a)$, $p \perp m$.
Therefore it follows from Claim~\ref{precispure} that $\preceq$ is
represented by a pure state, in  which case the claim holds.
\end{proof}
We now turn to the proof of the Theorem~\ref{triple}. Let $M$ be a
maximal element. If $\preceq$ is not trivial then  from the last
claim it follows that $M\perp u_i$ for every $i$. This is impossible
since since $u_1,u_2,u_3$ are linear independent, and the proof is
complete.

\begin{remark}
We do not know whether Theorem \ref{triple} holds true without the
assumption that $\preceq$ is continuous over one-dimensional spaces.
The proof hinges on this assumption in two ways. First, in that
$\preceq$ attains a minimum and a maximum. Second, in Claim
\ref{claim mmp} $x $ is chosen so that among all $x \sim m$,
$\theta(p,x)$ is minimal. While we could explicitly assume that
 $\preceq$ attains a minimum and a maximum, we could not dispose of
 the continuity assumption in the proof of Claim \ref{claim mmp}.
\end{remark}

%
\section{A Counterexample}\label{sec-counter}
In this section we present an example of a separable (though not
continuous) weak order over subspaces of $\mathbb{R}^3$ that
satisfies the standard assumptions but does not admit a
representation via a quantum measure. We need the following two
lemmas.
\begin{lem}
\label{counter_1} Let $\preceq$ be a weak order over one-dimensional
subspaces of $\mathbb{R}^3$ such that for every two-dimensional
subspace $U$ of $\mathbb{R}^3$ and every one-dimensional subspaces
$u,v$ of $U$ one has $u\preceq v\longleftrightarrow v'\preceq u'$
where $u',v'$ are the orthogonal complements of $u,v$ resp. in $U$,
then $\preceq$ can be extended to a weak order over $\mathbb{R}^3$
that satisfies the standard assumptions.
\end{lem}
\begin{proof}
We define $\preceq$ as follows. Let $U,V$ be two subspaces of
$\mathbb{R}^3$. If $\dim(U)<\dim(V)$ then $U\prec V$. If
$\dim(U)=\dim(V)=2$ then $U\preceq V$ iff $V^\perp\preceq U^\perp$.
\textbf{Negation} is obviously satisfied. As for
\textbf{de-Finetti's axiom}, let $u,v$ be two different
one-dimensional subspaces and $x$ be the one-dimensional subspaces
such that $x\perp u,v$. Let $u',v'$ be the orthogonal complements of
$u,v$ in $u+v$. Then $(x+u)^\perp=u'$ and $(x+v)^\perp=v'$. Since,
by the assumption of the lemma $v'\preceq u'$, it follows by
definition of $\preceq$ that $x+u\preceq x+v$.
\end{proof}

The second lemma states that if $\preceq$ is represented by a
probability measure, then the order over one-dimensional subspaces
of a fixed two-dimensional subspace $U$ has a very specific form. As
usual we identify one-dimensional subspaces with unit vectors. If
$S^1$ is the unit circle of $U$, the lemma essentially says that
either all elements of $S^1$ are equivalent, or there is a single
maximal element $x\in S^1$ that satisfies $y_1\preceq y_2$ iff $y_2$
is closer than $y_1$ to $x$.
\begin{lem}
\label{counter_2} Let $\mu$ be a probability measure over
$\mathbb{R}^3$ and $\preceq$ the corresponding weak order over
subspaces. Let $U$ be a two-dimensional subspace of $\mathbb{R}^3$.
Then, either all one-dimensional subspaces of $U$ are equivalent, or
there exists unit vector $x\in U$ such that for every pair $y_1,y_2$
of unit vectors $y_1\preceq y_2$ iff $|\inner{x,y_1}|\leq
|\inner{x,y_2}|$.
\end{lem}
\begin{proof}
By Gleason's Theorem, there exists a positive semidefinite operator
$T$ such that $\mu(A)=\tr(\Pi_A T)$. Consider the operator $\Pi_U
T\Pi_U$. This is a positive semidefinite operator. Its spectral
decomposition is of the form
\[
\Pi_U T\Pi_U=\alpha\Pi_x + \beta\Pi_{x'},\] where $x,x'$ are
orthogonal eigenvectors in $U$ with corresponding eigenvalues
$\alpha,\beta$ such that $\alpha+\beta=1$. We assume that
$\alpha\geq\beta$. It follows that for every unit vector $y$ in $U$,
\begin{multline*}
\mu(y)=\tr(\Pi_y T)=\tr(\pi_y \Pi_U T
\Pi_U)=\\=\alpha|\inner{y,x}|^2+\beta|\inner{y,x'}|^2=\beta+(\alpha-\beta)|\inner{y,x}^2|.\end{multline*}
Thus, if $\alpha=\beta$ then all $y\in S^2\cap U$ are equivalent. If
$\alpha>\beta$ then $\mu(y)$ is a monotonic function of
$|\inner{y,x}|$.  \end{proof}
\begin{example} \label{ex counter} Let $\succeq'$ be a weak order on one-dimensional subspaces of
$\mathbb{R}^2$ that satisfies the condition of Lemma~\ref{counter_1}
but not the condition of Lemma~\ref{counter_2}. Define $\preceq$ on
one-dimensional subspaces of $\mathbb{R}^3$ as follows: Let $p$ be
the northern pole of the unit sphere in $\mathbb{R}^3$. Let $u$ and
$v$ be unit vectors, then $u\succeq v$ either when $u\not \in E$ and
$\mu(u) \ge \mu(v)$ or when $u,v \in E$ and $u \succeq' v$. By
Lemma~\ref{counter_1} $\prec$ can be extended to a weak order over
$\mathbb{R}^3$ that satisfies the standard assumptions. However,
since the condition of Lemma~\ref{counter_2} is not satisfied,
$\preceq$ cannot be represented via a quantum measure.
\end{example}

\section{Final Comments and Open Problems}\label{sec-final}
\subsection{Representation and continuity}

 In Gleason's Theorem (\cite{gleason}) continuity is not assumed and
is a consequence of the existence of a frame function. When the
primitive of the model is a likelihood order, matters are different.
The likelihood order in Example \ref{ex counter} satisfies
\textbf{de-Finetti's axiom}, \textbf{Negation},
\textbf{Monotonicity} and \textbf{Separability} and cannot be
represented by a quantum measure. This order, which is not
continuous, suggests that continuity must be explicitly assumed and
cannot be derived from more plausible assumptions.

The question whether every continuous likelihood order which
satisfies \textbf{de-Finetti's axiom}, \textbf{Negation},
\textbf{Monotonicity} and \textbf{Separability} can be represented
by a quantum measure is still open.

\subsection{Partial representation}
\begin{defn}
 We say that $\mu$ \emph{partially represents}
  $\preceq$ if $A\preceq B\longrightarrow \mu(A)\leq \mu(B)$ for
  every two subspaces $A,B$ of $\calh$.\end{defn}

It turns out (we state without a proof) that if $\preceq$ satisfies
the cancelation condition, then there exists a quantum probability
measure that partially represents $\preceq$. Also, from the proof of
Theorem~\ref{purethm} it follows that, if there exists a
one-dimensional subspace $p$, such that $p\sim \calh$, then (without
assuming separability) $\preceq$ admits a partial representation by
a pure state.

%

\subsection{Qualitative additivity  and discrete orders}

Gleason's Theorem implies that the only quantum probability measure
which obtains a discrete set of values is the uniform measure. The
question arises whether the same is true for likelihood orders. We
say that $\preceq$ is \emph{discrete} if the restriction of $\sim$
to one-dimensional subspaces has only finitely many equivalence
classes. For instance,  if $\preceq$ is represented by the uniform
probability, then its restriction to one-dimensional subspaces has
only one equivalence class.

Kochen-Specker's Theorem (\cite{K-S}) actually refers to likelihood
orders whose restriction to one-dimensional subspaces have precisely
two equivalence classes. In order to prove this result using
likelihood orders terms only, one needs to strengthen
\textbf{de-Finetti's axiom} are \textbf{Negation}. The following
axiom is  a consequence of \textbf{de-Finetti's axiom} in the
classical case, but not in the a quantum set-up.

\bigskip

\textbf{Qualitative additivity}: Let $A_1,A_2,B_1,B_2$ be  linear
subspaces of $\calh$ such that $A_1\perp A_2$ and  $B_1\perp B_2$.
If $A_i\preceq B_i$, $i=1,2$ then $A_1\oplus A_2\preceq B_1\oplus
B_2$. Furthermore, one strict likelihood on the former inequalities
implies strict likelihood in the later inequality.

\bigskip

Suppose that $A_1$ and $A_2$ are orthogonal and the same for the
$B_i$'s. \textbf{Qualitative additivity} states that
, if the $A_i$'s are at least as likely as $B_i$'s, then the
subspace spanned by the $A_i$'s is at least as likely as that
spanned by $B_i$'s. That is, adding or subtracting a more likely
subspace to or from a subspace which is already more likely, cannot
make the outcome less likely.

Suppose that $\preceq$ is defined over $\mathbb{R}^3$ and there are
only two equivalence classes of one-dimensional subspaces, say,
green and red. If $\preceq$ satisfies \textbf{Qualitative
additivity}, then in any orthogonal triple there is the same number
of green representatives, and  moreover, a two dimensional subspace
spanned by uni-colored vectors contains only vectors of the same
color. These are precisely the terms of Kochen-Specker's Theorem
(\cite{K-S}). It states that there exists no likelihood order that
satisfies  \textbf{Qualitative additivity}  and has precisely two
one-dimensional equivalence classes.

This result suggests that the only  discrete likelihood order that
satisfies \textbf{Qualitative additivity} is that induced by the
uniform measure.

\end{document}